\documentclass[conference]{IEEEtran}
\IEEEoverridecommandlockouts
\usepackage{cite}
\usepackage{amsmath,amssymb,amsfonts}
\usepackage{algorithm}
\usepackage{algorithmic}
\usepackage{graphicx}
\usepackage{epstopdf}
\usepackage{booktabs}
\usepackage{makecell}
\usepackage{multirow}
\usepackage{textcomp}
\usepackage{xcolor}
\usepackage[colorlinks  = true,
            linkcolor   = black,
            urlcolor    = magenta,
            anchorcolor = blue,
            bookmarks   = false
            ]{hyperref}

\def\BibTeX{{\rm B\kern-.05em{\sc i\kern-.025em b}\kern-.08em
    T\kern-.1667em\lower.7ex\hbox{E}\kern-.125emX}}
\graphicspath{ {./images/} }
\DeclareMathOperator*{\argmin}{arg\,min}
\DeclareMathOperator*{\argmax}{arg\,max}

\begin{document}

\title{Binary Neural Network Aided CSI Feedback in Massive MIMO System \\
}

\author{\IEEEauthorblockN{Zhilin Lu}
\IEEEauthorblockA{\textit{Beijing National Research Center for}\\ \textit{Information Science and Technology}\\ \textit{(BNRist), Tsinghua University}\\
Beijing, China \\
luzl18@mails.tsinghua.edu.cn}\and

\IEEEauthorblockN{Jintao Wang}
\IEEEauthorblockA{\textit{Beijing National Research Center for}\\ \textit{Information Science and Technology}\\ \textit{(BNRist), Tsinghua University}\\
Beijing, China \\
wangjintao@tsinghua.edu.cn}\and

\IEEEauthorblockN{Jian Song}
\IEEEauthorblockA{\textit{Beijing National Research Center for}\\ \textit{Information Science and Technology}\\ \textit{(BNRist), Tsinghua University}\\
Beijing, China \\
jsong@tsinghua.edu.cn}\and
}
\maketitle

\begin{abstract}
In massive multiple-input multiple-output (MIMO) system, channel state information (CSI) is essential for the base station to achieve high performance gain. Recently, deep learning is widely used in CSI compression to fight against the growing feedback overhead brought by massive MIMO in frequency division duplexing system. However, applying neural network brings extra memory and computation cost, which is non-negligible especially for the resource limited user equipment (UE). In this paper, a novel binarization aided feedback network named BCsiNet is introduced. Moreover, BCsiNet variants are designed to boost the performance under customized training and inference schemes. Experiments shows that BCsiNet offers over 30\(\times\) memory saving and around 2\(\times\) inference acceleration for encoder at UE compared with CsiNet. Furthermore, the feedback performance of BCsiNet is comparable with original CsiNet. The key results can be reproduced with \textnormal{\href{https://github.com/Kylin9511/BCsiNet}{https://github.com/Kylin9511/BCsiNet}}.
\end{abstract}


\section{Introduction}
Massive multiple-input multiple-output (MIMO) is a promising technique for 5G wireless communication systems, providing better the spectrum and energy efficiency \cite{larsson2014massive}\cite{boccardi2014five}. However, the base station (BS) needs realtime channel state information (CSI) to acquire performance gain. In frequency division duplexing (FDD) system, downlink CSI must be fed back from user equipment (UE) due to the asymmetry of uplink and downlink channel. With the growing numbers of transmitting antennas in massive MIMO system, the feedback overhead of the CSI matrix becomes unbearable.

Recently, deep learning (DL) is widely adopted to wireless communication scenarios including downlink CSI feedback. Generally, the CSI matrix is compressed by a neural network (NN) based encoder at UE to reduce feedback overhead. CsiNet\cite{wen2018deep} is the first to prove the effectiveness of the DL based CSI feedback scheme over traditional compressed sensing (CS) algorithms \cite{kuo2012compressive}.

Many works make remarkable contributions by extending the basic scenario proposed in \cite{wen2018deep}. Time-varying CSI is considered in \cite{wang2018deep} while correlation between uplink and downlink CSI is utilized in \cite{liu2019exploiting, yang2019deep}. Entropy quantizer is adopted in \cite{yang2019deepcmc}. A denoise module is added to deal with imperfect feedback in \cite{ye2020deep, sun2020ancinet}. At the mean while, series of papers introduce novel network design for performance boosting. CRNet\cite{lu2020multi} and CsiNetPlus\cite{guo2020convolutional} improve the feedback capacity with multi-resolution network and network expansion, respectively. Pseudo-3D convolution is utilized in \cite{li2020spatio} and fully connected (FC) layers is removed in \cite{cao2020lightweight}.  Squeeze and excitation network is used in \cite{cai2019attention} while non-local block is applied in \cite{yu2020ds}.

However, most of these works trade higher performance with extra memory and computation cost. This can be impractical for many user equipments whose hardware resources are strictly limited. Existing papers devoted to lightweight network design mainly based on larger network like CsiNetPlus\cite{guo2020convolutional} and ConvCsiNet \cite{cao2020lightweight}. Therefore many of the final lightweight networks are actually heavier than CsiNet in computation complexity or parameters size. \cite{guo2020compression} introduces FC pruning and network quantization and gives impressive results. However, pruning with unstructured weights brings difficulty to inference at UE and vanilla model quantization harms the performance deeply when the quantized bits is lower than five.

In this paper, we propose a novel lightweight feedback network named binary CsiNet (BCsiNet). Following basic principles proposed in \cite{courbariaux2015binaryconnect, rastegari2016xnor}, we successfully binarize the FC layer at UE. Moreover, we expand the BCsiNet design to boost the performance without much extra cost. Simulation shows that BCsiNet has comparable performance against the original CsiNet while offering over $30\times$ memory saving and around $2\times$ acceleration at UE.

The main contributions of this paper is listed as follows.
\begin{itemize}
    \item Binary neural network (BNN) technique is introduced to CSI feedback task and proved to be effective. To the author's best knowledge, we are the first to apply BNN into extremely lightweight feedback encoder design.
    \item Neat and valid manner is given to generate BCsiNet variants. The expanded BCsiNet has better performance with little extra cost.
    \item Special training and inference schemes are designed for BCsiNet, which is essential for training convergence and inference acceleration.
\end{itemize}

The rest of the paper is structured as follows. System model is proposed in section \ref{Section-SystemModel} while detailed design of BCsiNet is given in section \ref{Section-BNN}. Section \ref{Section-Results} presents the numerical results and analysis. Section \ref{Section-Conclusion} concludes the paper in the end.

\begin{figure*}[t]
\centering
\includegraphics[width=\textwidth]{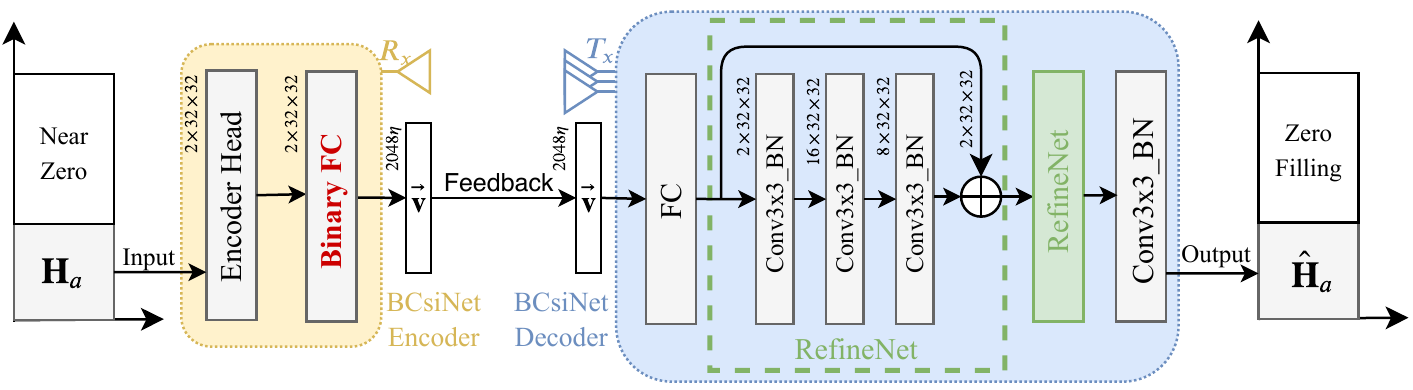}
\caption{Overview of BCsiNet aided downlink CSI feedback workflow. The fully connected layer of the encoder is binarized.}
\label{System Model}
\end{figure*}

\section{System Model} \label{Section-SystemModel}

We consider a single-cell massive MIMO FDD system with $N_c$ orthogonal frequency division multiplexing (OFDM) sub-carriers. There are $N_t$($N_t \gg 1$) transmitting antennas at BS and $N_r$ receiving antennas at UE. We take $N_r = 1$ for simplicity. The received signal can be expressed as follows:

\begin{equation}
    \mathbf{y} = \mathbf{A}\mathbf{x} + \mathbf{z},
\end{equation}

where $\mathbf{x},\mathbf{y},\mathbf{z} \in \mathbb{C}^{N_c\times 1}$ are the transmitted signal, received signal and additive noise in one OFDM period. $\mathbf{A} = \text{diag}(\tilde{\mathbf{h}}_1^H\mathbf{p}_1, \cdots, \tilde{\mathbf{h}}_{N_c}^H\mathbf{p}_{N_c})$ is a diagonal matrix, where $\tilde{\mathbf{h}}_i, \mathbf{p}_i \in \mathbb{C}^{N_t\times 1}, i \in \{1, \cdots, N_c\}$ are the downlink channel vector and precoding vector at sub-carrier $i$, respectively.

The base station needs to obtain the channel vector $\tilde{\mathbf{h}}_i$ in order to design the precoding vector $\mathbf{p}_i$, for which all the $\tilde{\mathbf{h}}_i$ must be fed back from UE. We define the overall downlink channel matrix as $\tilde{\mathbf{H}} \triangleq [\tilde{\mathbf{h}}_1, \cdots, \tilde{\mathbf{h}}_{N_c}]^H$, which contains $2N_cN_t$ elements.

The overhead is too large to directly feed all $2N_cN_t$ elements back. Like it is shown in \cite{wen2018deep}, the channel matrix is sparse in the angular-delay domain. We can take advantage of it and transfer the CSI matrix from spatial-frequency domain to angular-delay domain via discrete Fourier transform (DFT).

\begin{equation} \label{eq2}
    \mathbf{H} = \mathbf{F}_c \tilde{\mathbf{H}} \mathbf{F}_t^H,
\end{equation}

where $F_c$ and $F_t$ are $N_c\times N_c$ and $N_t\times N_t$ DFT matrices, respectively. Since the time delay of multi-path arrivals is limited, only the first $N_a$ rows of CSI matrix $\mathbf{H}$ contains large components. The rest of the rows with near-zero elements can be truncated. We denote the first $N_a$ rows of the CSI matrix with $\mathbf{H}_a$ for simplicity.

However, the matrix $\mathbf{H}_a$ is still too heavy in massive MIMO system where $N_t$ is large. That is why $\mathbf{H}_a$ needs to be further compressed before feedback. Traditional compressed sensing algorithms require $\mathbf{H}_a$ to be sparse enough, but the sparsity of $\mathbf{H}_a$ is guaranteed only when $N_t \rightarrow \infty$ which is impractical \cite{wen2014channel}. Neural network based encoder-decoder can work without such limitation and achieve a better CSI compressing and reconstructing performance.

The overall feedback workflow is demonstrated in Fig. \ref{System Model}. At the beginning, the BCsiNet encoder takes the truncated CSI matrix $\mathbf{H}_a$ as input at UE. The compressed channel feature $\mathbf{v}$ is generated and fed back to BS. Then the BCsiNet decoder reconstruct $\mathbf{H}_a$  and the original CSI matrix $\tilde{\mathbf{H}}$ is achieved by zero filling and inverse DFT. We can summarize the aforementioned feedback workflow with equation (\ref{eq3}).

\begin{equation} \label{eq3}
    \hat{\mathbf{H}}_a = F_d(F_e(\mathbf{H}_a, \Theta_e), \Theta_d),
\end{equation}

where $F_e$ represents BCsiNet encoder with parameters $\Theta_e$ and $F_d$ represents BCsiNet decoder with parameters $\Theta_d$. With $\Theta_e$ and $\Theta_d$ properly optimized, distance between $\mathbf{H}_a$ and $\hat{\mathbf{H}}_a$ can be minimized. Our purpose here is to lighten $\Theta_e$ using FC binarization to reduce the memory and power consumption together with computational complexity.

It is worth mentioning that the uplink feedback is assumed to be ideal in this paper. Besides, the COST2100 \cite{liu2012cost} channel model is used to simulate the CSI matrix.

\section{Binary Neural Network in CSI Feedback} \label{Section-BNN}

\subsection{Complexity Analysis of Existing Feedback Networks} \label{SubSection-ComplexityAnalysis}
Before explaining how binary neural network works, we give a glance at complexity of several existing feedback networks to provide a contrastive point of view.

As it is presented in Table \ref{tab1}, later feedback networks mainly focus on trading better performance with larger cost. Therefore, CsiNet actually has the lightest encoder among all these networks. By further compressing the CsiNet encoder, we are able to provide an extremely lightweight encoder that can works under much tougher resource restriction at UE.

In order to further compress the CsiNet encoder, we need to analyze the distribution of its complexity. As we can see in Table \ref{tab1} and Table \ref{tab2}, the complexity proportion of FC goes higher as the encoder gets simpler. With simplest encoder, FC layer takes up 96.61\% and 99.996\% of computation and memory resources of original CsiNet, respectively.

From the aforementioned observation, it is clear that a lighter fully connected layer is the key to the extremely lightweight CsiNet encoder design. Note that the flops counting in Table \ref{tab1} and Table \ref{tab2} ignores the batch normalization (BN) layer. BN is cost free in inference since it can be merged into the corresponding convolution layer.

\begin{table}[t]
\caption{params and flops of several existing feedback networks}
\begin{center}
\begin{tabular}{l|c c|c c}
\Xhline{0.8pt}
\multirow{2}{*}{\textbf{Methods}$^{\mathrm{a}}$} & \multicolumn{2}{c|}{\textbf{Encoder at UE}} & \multicolumn{2}{c}{\textbf{Decoder at BS}}\\
& FLOPs$^{\mathrm{b}}$ & params & FLOPs$^{\mathrm{b}}$ & params \\
\Xhline{0.8pt}
CsiNet \cite{wen2018deep} & 1.09M & 1.05M & 4.33M & 1.05M \\
CRNet \cite{lu2020multi} & 1.20M & 1.05M & 3.92M & 1.05M \\
CsiNetPlus \cite{guo2020convolutional} & 1.45M & 1.05M & 23.12M & 1.07M \\
ConvCsiNet \cite{cao2020lightweight} & 60.16M & 2.14M & 166.07M & 2.07M \\
DeepCMC \cite{yang2019deepcmc} & 173.54M & 3.32M & 278.40M & 9.87M \\
\Xhline{0.8pt}
\multicolumn{5}{l}{$^{\mathrm{a}}$ The compression ratio $\eta$ is $1/4$ for all methods.} \\
\multicolumn{5}{l}{$^{\mathrm{b}}$ FLOPs is total number of "multiply then add" operation.} \\
\end{tabular}
\label{tab1}
\end{center}
\end{table}

\begin{table}[t]
\caption{encoder complexity distribution of feedback networks}
\begin{center}
\begin{tabular}{l|c c|c c}
\Xhline{0.8pt}
\multirow{2}{*}{\textbf{Methods}$^{\mathrm{a}}$} & \multicolumn{2}{c|}{\textbf{\;\;\;\;\;\; UE FLOPs$^{\mathrm{b}}$\;\;\;\;\;\;}} & \multicolumn{2}{c}{\textbf{\;\;\;\;\;\; UE params\;\;\;\;\;\;}}\\
& FC & others$^{\mathrm{c}}$ & FC & others$^{\mathrm{c}}$ \\
\Xhline{0.8pt}
CsiNet \cite{wen2018deep} & 96.60\% & 3.40\% & 99.996\%  & 0.004\% \\
CRNet \cite{lu2020multi} & 87.07\% & 12.93\% & 99.984\% & 0.016\% \\
CsiNetPlus \cite{guo2020convolutional} & 72.32\% & 27.68\% & 99.962\% & 0.038\% \\
\Xhline{0.8pt}
\multicolumn{5}{l}{$^{\mathrm{a}}$ The compression ratio $\eta$ is $1/4$ for all methods.} \\
\multicolumn{5}{l}{$^{\mathrm{b}}$ FLOPs is the total number of "multiply then add" operation.} \\
\multicolumn{5}{l}{$^{\mathrm{c}}$ all other layers including convolution, batch normalization, etc.} \\
\end{tabular}
\label{tab2}
\end{center}
\end{table}

\subsection{Binarization Algorithm Design for Fully Connected Layer} \label{SubSection-BinarizationAlgorithm}

In this subsection, we will explain how to slim the FC layer at UE with network binarization. Compared with standard NN, BNN is able to save around $32\times$ memory since the parameters change from 32bits float point numbers to 1bit binary numbers.

Moreover, the ``multiply-accumulate'' operation in standard NN is replaced by simple addition in BNN as it is shown in equation (\ref{eq4}). Generally in one ``multiply-accumulate'' operation, multiplication takes more time than addition on chips. For instance, the latency of float multiplication and addition is 4 and 2 in VIA Nano 2000 series of CPU \cite{chen2020addernet}. Therefore, the speed of BNN inference is over $2\times$ faster than standard NN since the time consuming multiplications are gone.

\begin{equation} \label{eq4}
\begin{aligned}
    &
    \left[\begin{matrix}
        \alpha_1 & \alpha_2 \\
        \alpha_3 & \alpha_4 \\
    \end{matrix}\right]
    \cdot
    \left[\begin{matrix}
        \beta_1 \\
        \beta_2 \\
    \end{matrix}\right]
    =
    \left[\begin{matrix}
        \alpha_1\times\beta_1 + \alpha_2\times\beta_2 \\
        \alpha_3\times\beta_1 + \beta_4\times\beta_2 \\
    \end{matrix}\right] && \text{for}\;\;\text{NN}\\
    &
    \left[\begin{matrix}
        1 & -1 \\
        1 & 1 \\
    \end{matrix}\right]
    \cdot
    \left[\begin{matrix}
        \beta_1 \\
        \beta_2 \\
    \end{matrix}\right]
    =
    \left[\begin{matrix}
        \beta_1 - \beta_2 \\
        \beta_1 + \beta_2 \\
    \end{matrix}\right] && \text{for}\;\;\text{BNN}\\
\end{aligned}
\end{equation}

However, binarization of the FC layer may harm the network performance. In order to reduce the information loss of binarization, a positive scale factor $\alpha \in \mathbb{R}^+$ is attached to the binary FC layer. For standard FC weight $\mathbf{W} \in \mathbb{R}^{m\times n}$ and binary FC weight $\mathbf{B} \in \{-1, 1\}^{m\times n}$, the distance between $\mathbf{W}$ and  $\mathbf{W}_b=\alpha\mathbf{B}$ should to be minimized. Denoting the vectorized matrix as $\mathbf{w} \triangleq \text{vec}(\mathbf{W}) \in \mathbb{R}^{mn}$ and $\mathbf{b} \triangleq \text{vec}(\mathbf{B}) \in \{-1, 1\}^{mn}$, we can formulate the optimization problem as follows.

\begin{equation} \label{eq5}
    (\mathbf{b}^{opt}, \alpha^{opt}) = \argmin_{\mathbf{b}, \alpha}\Vert \mathbf{w} - \alpha\mathbf{b}\Vert^2
\end{equation}

Following the derivation in \cite{rastegari2016xnor}, we expand the original distance in (\ref{eq5}).

\begin{equation} \label{eq6}
\begin{aligned}
    \Vert \mathbf{w} - \alpha\mathbf{b}\Vert^2
    &= \left(\mathbf{b}^T\mathbf{b}\alpha^2 - 2\mathbf{w}^T\mathbf{b}\alpha +  \mathbf{w}^T\mathbf{w}\right) \\
    &= \left(mn\alpha^2 - 2\mathbf{w}^T\mathbf{b}\alpha +  \mathbf{w}^T\mathbf{w}\right) \\
    &\triangleq \left(mn\alpha^2 - 2\mathbf{w}^T\mathbf{b}\alpha + Z\right)_, \\
\end{aligned}
\end{equation}

where $Z = \mathbf{w}^T\mathbf{w}$ is a const since $\mathbf{w}$ is given. It is obvious that $\mathbf{w}^T\mathbf{b}$ should be maximized to minimize equation (\ref{eq6}).

\begin{equation} \label{eq7}
    \mathbf{b}^{opt} = \argmax_{\mathbf{b}}\mathbf{w}^T\mathbf{b} = \text{sign}(\mathbf{w})
\end{equation}

Then $\alpha^{opt}$ can be deduced from the derivative of (\ref{eq6}).

\begin{equation} \label{eq8}
\begin{aligned}
    \alpha^{opt}
    &= -\frac{-2\mathbf{w}^T\mathbf{b}^{opt}}{2mn} \\
    &= \frac{\mathbf{w}^T\text{sign}(\mathbf{w})}{mn} \\
    &= \frac{1}{mn}\sum_{i=1}^{mn}\vert \mathbf{w}_i\vert = \frac{1}{mn}\Vert\mathbf{W}\Vert_1 \\
\end{aligned}
\end{equation}

\begin{figure}[b]
\centering
\includegraphics[width=0.48\textwidth]{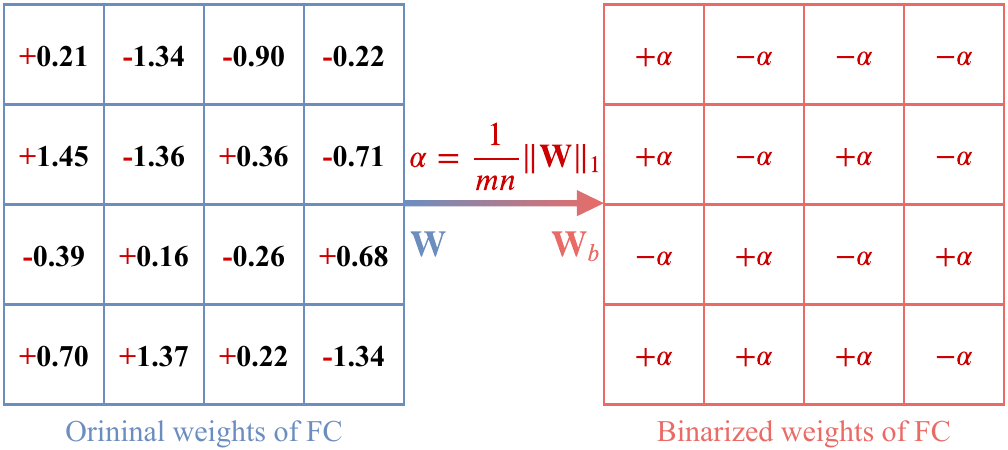}
\caption{A demonstration of fully connected layer binarization with scale.}
\label{FC Binary}
\end{figure}

Therefore the final binarized $\mathbf{W}_b$ can be derived as follows.

\begin{equation} \label{eq9}
    \mathbf{W}_b = \alpha\cdot\text{sign}(\mathbf{W}) = \frac{1}{mn}\Vert\mathbf{W}\Vert_1\text{sign}(\mathbf{W})
\end{equation}

A demonstration of the FC binarization with scale is given in Fig. \ref{FC Binary} for better intuitive understanding. It is worth mentioning that we only binarize the weight in FC layer while the bias remains unchanged. This enhances the capacity of Binary FC layers with little cost.

\subsection{Design of the proposed BCsiNet} \label{SubSection-DesignOfBCsiNet}
Vanilla BCsiNet can be obtained by directly applying binarization algorithm in section \ref{SubSection-BinarizationAlgorithm} to the encoder of CsiNet. As it is shown in Fig. \ref{System Model}, the encoder of BCsiNet is made up of an encoder head and a binary FC layer while the decoder consists of two RefineNets and a $3\times 3$ convolution layer.

\begin{figure}[t]
\centering
\includegraphics[width=0.48\textwidth]{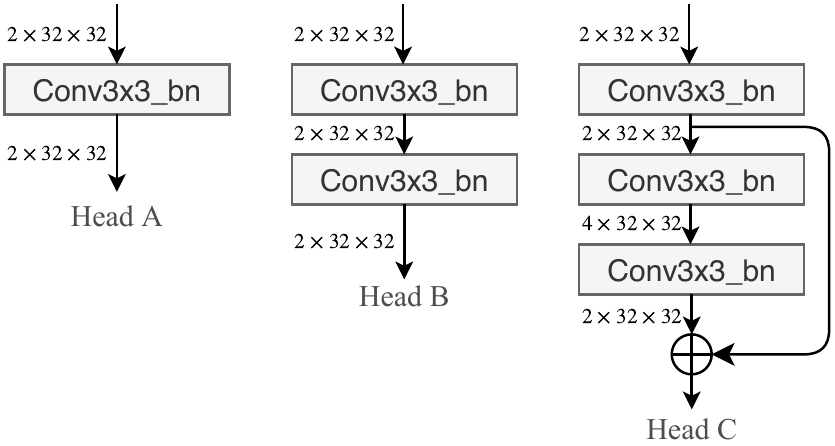}
\caption{Design of BCsiNet encoder heads.}
\label{Encoder Heads}
\end{figure}

In order to enhance the performance of proposed BCsiNet, we design several variants for it. The original encoder head of CsiNet is called ``head A'', which is single $3\times 3$ convolution layer. Head B and head C are designed to replace head A for better performance. As we can see in Fig. \ref{Encoder Heads}, head B adds a concatenated $3\times 3$ convolution layer to head A, while head C includes an extra residual architecture. Deeper encoder head like head B or head C can extract the channel feature better since the available resolution is larger.

In fact, extra cost of head B or head C is tiny and BCsiNet encoder is extremely light with all these heads. Compared with original CsiNet encoder, BCsiNet encoder saves $31.49\times$, $31.48\times$ and $31.34\times$ memory with head A, B and C, respectively. Similarly, applying different heads has little impact on the $2\times$ inference acceleration of BCsiNet.

On the other hand, we expand the original CsiNet decoder by adding one more RefineNet. Three concatenated RefineNets work better without extra cost at UE. The computational complexity rises from 4.32M to 5.95M with negligible parameter size increase, which is completely acceptable for the base station.

Note that all the convolution layers in BCsiNet are followed by BN layer and activation layer. LeakyReLU with negative slope of 0.3 is used as activation function. Specially, a sigmoid function replaces the original LeakyReLU at the end of BCsiNet decoder to limit the output range.

\subsection{Training and Inference of BCsiNet} \label{SubSection-TrainingAndInference}

Special training and inference scheme must be designed in order to train the proposed BCsiNet variants well. The most obvious problem is that the sign function introduced in binary FC layer is not derivable. In order to train an end-to-end network, we use a gate filter as \cite{rastegari2016xnor} to calculate the gradient of sign function. The gate filter is shown in equation (\ref{eq10}).

\begin{equation} \label{eq10}
    \frac{\partial \text{sign}(x)}{\partial x} =
    \left\{\begin{aligned}
        & x && \text{if}\;\; \vert x \vert < 1 \\
        & 0 && \text{otherwise} \\
    \end{aligned}\right.
\end{equation}

Based on (\ref{eq10}), the gradient of the scaled sign function in (\ref{eq9}) is derived as follows.

\begin{equation} \label{eq11}
\begin{aligned}
    \frac{\partial C}{\partial \mathbf{W}}
    &= \frac{\partial C}{\partial \mathbf{W}_b}\cdot\frac{\partial \mathbf{W}_b}{\partial \mathbf{W}} \\
    &= \frac{\partial C}{\partial \mathbf{W}_b}\cdot\frac{\partial \left(\alpha\cdot\text{sign}(\mathbf{W})\right)}{\partial \mathbf{W}} \\
    &= \frac{\partial C}{\partial \mathbf{W}_b}\left(\frac{1}{mn}+\alpha\frac{\partial \text{sign}(\mathbf{W})}{\partial \mathbf{W} }\right)_, \\
\end{aligned}
\end{equation}

where $C$ is the cost function of the network, and $\frac{\partial C}{\partial \mathbf{W}}$ is the gradient of the encoder FC layer in back-propagation.

It is also important to notice that all the parameter updates are based on float point weights. Typically, the parameters change in a single iteration is so small that the weights remain the same if they are binarized. Nothing can be learned if the weights are not updated. Therefore, float point weights are essential during training for network convergency. Note that the float point parameters of binary FC layer do no harm to inference since they only exist during training. The detailed training pipeline of BCsiNet is listed in Algorithm \ref{algA}.

\begin{algorithm}[!b]
\renewcommand{\algorithmicrequire}{\textbf{Input:}}
\renewcommand\algorithmicensure {\textbf{Output:}}
\caption{A training iteration of proposed BCsiNet}
\label{algA}
\begin{algorithmic}[1]

\REQUIRE Input CSI data $\mathbf{D}^t$ of current mini-batch, current learning rate $\gamma^t$, current weights of BCsiNet $\mathcal{W}^t$.
\ENSURE Updated weights of BCsiNet $\mathcal{W}^{t+1}$ and updated learning rate $\gamma^{t+1}$.
\STATE Save the real-value weight of encoder FC layer. $\tilde{\mathbf{W}^t} = \mathbf{W}^t$.
\STATE Calculate the binary scale $\alpha$ and the binarized weight $\mathbf{W}_b^t$ of encoder FC layer according to (\ref{eq9}).
\STATE Assign the encoder FC layer with the binarized weight $\mathbf{W}^t = \mathbf{W}_b^t$, making the forward propagation binary.
\STATE Execute the forward and backward propagation, getting the gradient of the binarized encoder FC layer $\frac{\partial C}{\partial \mathbf{W}_b^t}$.
\STATE Restore the original real-value weight of encoder FC layer for parameter update $\mathbf{W}^t = \tilde{\mathbf{W}^t}$.
\STATE Calculate the real-value weight gradient of the encoder FC layer $\frac{\partial C}{\partial \mathbf{W}^t}$ based on (\ref{eq11}).
\STATE Update all the parameters in BCsiNet with Adam optimizer. $\mathcal{W}^{t+1} = \textbf{AdamOptimizerStep}(\mathcal{W}^t, \frac{\partial C}{\partial \mathcal{W}^t}, \gamma^t)$.
\STATE Update the learning rate. $\gamma^{t+1}=\textbf{LRScheduler}(\gamma^t, t)$.

\end{algorithmic}
\end{algorithm}

When training completes, the binarized FC weight $\mathbf{B} = \text{sign}(\mathbf{W})$ and the attached scale $\alpha = \frac{1}{mn}\Vert \mathbf{W} \Vert_1$ are calculated and saved for inference. Note that $\alpha$ is a float point number. Therefore, the input feature should be multiplied with the binary weight $\mathbf{B}$ in advance for higher inference efficiency. The correct computing order is shown in equation (\ref{eq12}).

\begin{equation} \label{eq12}
    \mathbf{x}_{i+1} = \mathbf{W}_b\mathbf{x}_i + \mathbf{b} = \alpha\mathbf{B}\mathbf{x}_i + \mathbf{b} = \alpha(\mathbf{B}\mathbf{x}_i) + \mathbf{b},
\end{equation}

where $\mathbf{x}_i$, $\mathbf{x}_{i+1}$ and $\mathbf{b}$ are input feature, output feature and bias of the binary FC layer, respectively.

\section{Simulation Results and Analysis} \label{Section-Results}

\subsection{Experiment Settings}

The proposed BCsiNet is trained with two different channel scenarios. The indoor scenario works under 5.3GHz band while the outdoor scenario works under 300MHz band. For fair comparison, all the parameter settings of COST2100 channel model are inherited from CsiNet \cite{wen2018deep}. The base station adopts a uniform linear array (ULA) system with $N_t=32$. A FDD system with $N_c=1024$ is considered and the angular-delay domain matrix is truncated to $N_a=32$. The training, validation and test dataset are independently generated with 100,000, 30,000 and 20,000 channel matrices, respectively.

Xavier initialization is used for both convolution and fully connected layers. Adam optimizer with $\beta_1=0.9$, $\beta_2=0.999$ and $\varepsilon=1e-7$ is adopted while the mean square error (MSE) loss is applied. The batch size is set to 1000 for faster training. The scheduler proposed in CRNet \cite{lu2020multi} is used as (\ref{eq13}).

\begin{equation} \label{eq13}
    \gamma = \gamma_{ed} + \frac{1}{2}(\gamma_{st} - \gamma_{ed})\left(1+\cos \left(\frac{i-N_w}{N-N_w}\pi\right)\right)_,
\end{equation}

where $N$, $N_w$, $i$ are number of total training epochs, number of warm up epochs and index of current epoch, respectively. Initial lr $\gamma_{st}$ is $1e-2$ while the final lr $\gamma_{ed}$ is $5e-5$. For BCsiNet training, we set $N=2500$ and $N_w=30$. A linear increasing scheduler is applied for warm up. Note that the training of BNN is trapped into bad local minima occasionally, and the problem can be settled by scheduler rebooting. All the training and inference are based on PyTorch.

\subsection{Performance and Complexity of the Proposed BCsiNet}

In order to evaluate the performance of CSI reconstruction, normalized mean square error (NMSE) is used to measure the distance between the original $\mathbf{H}_a$ and the recovered $\hat{\mathbf{H}}_a$.

\begin{equation} \label{eq14}
    \text{NMSE} = \mathsf{E}\left\{\frac{\Vert\mathbf{H}_a-\hat{\mathbf{H}}_a\Vert_2^2}{\Vert\mathbf{H}_a\Vert_2^2}\right\}
\end{equation}

As it is explained in section \ref{SubSection-DesignOfBCsiNet}, we design several variants for the proposed BCsiNet. Table \ref{tab3} shows an ablation study among different BCsiNet variants under indoor and outdoor scenario. The compression ratio $\eta$ is fixed to $1/4$.

\begin{table}[!b]
\caption{NMSE (dB) comparison among BCsiNet variants}
\begin{center}
\begin{tabular}{c|c c|c c}
\Xhline{0.8pt}
\multirow{2}{*}{\textbf{BCsiNet Head}} & \multicolumn{2}{c|}{\textbf{Two RefineNets}} & \multicolumn{2}{c}{\textbf{Three RefineNets}}\\
& Indoor & Outdoor & Indoor & Outdoor \\
\Xhline{0.8pt}
A & -17.25 & -8.35 & -17.49 & -8.78 \\
B & \textbf{-19.00} & -9.07 & \textbf{-20.31} & -9.77 \\
C & -18.32 & \textbf{-9.20} & -19.00 & \textbf{-9.93} \\
\Xhline{0.8pt}
\end{tabular}
\label{tab3}
\end{center}
\end{table}

As we can see from Table \ref{tab3}, the performance of BCsiNet-B and BCsiNet-C is higher than vanilla BCsiNet-A under both indoor and outdoor scenario. Moreover, it is clear that B is relatively dominant for indoor scenario while C works better under outdoor scenario. Besides, ablation study in Table \ref{tab3} proves that NMSE performance can be further improved by extending an extra RefineNet block at decoder. The extension is cost free for UE and acceptable for BS like it is mentioned in section \ref{SubSection-DesignOfBCsiNet}.

\begin{figure}[t]
\centering
\includegraphics[width=0.48\textwidth]{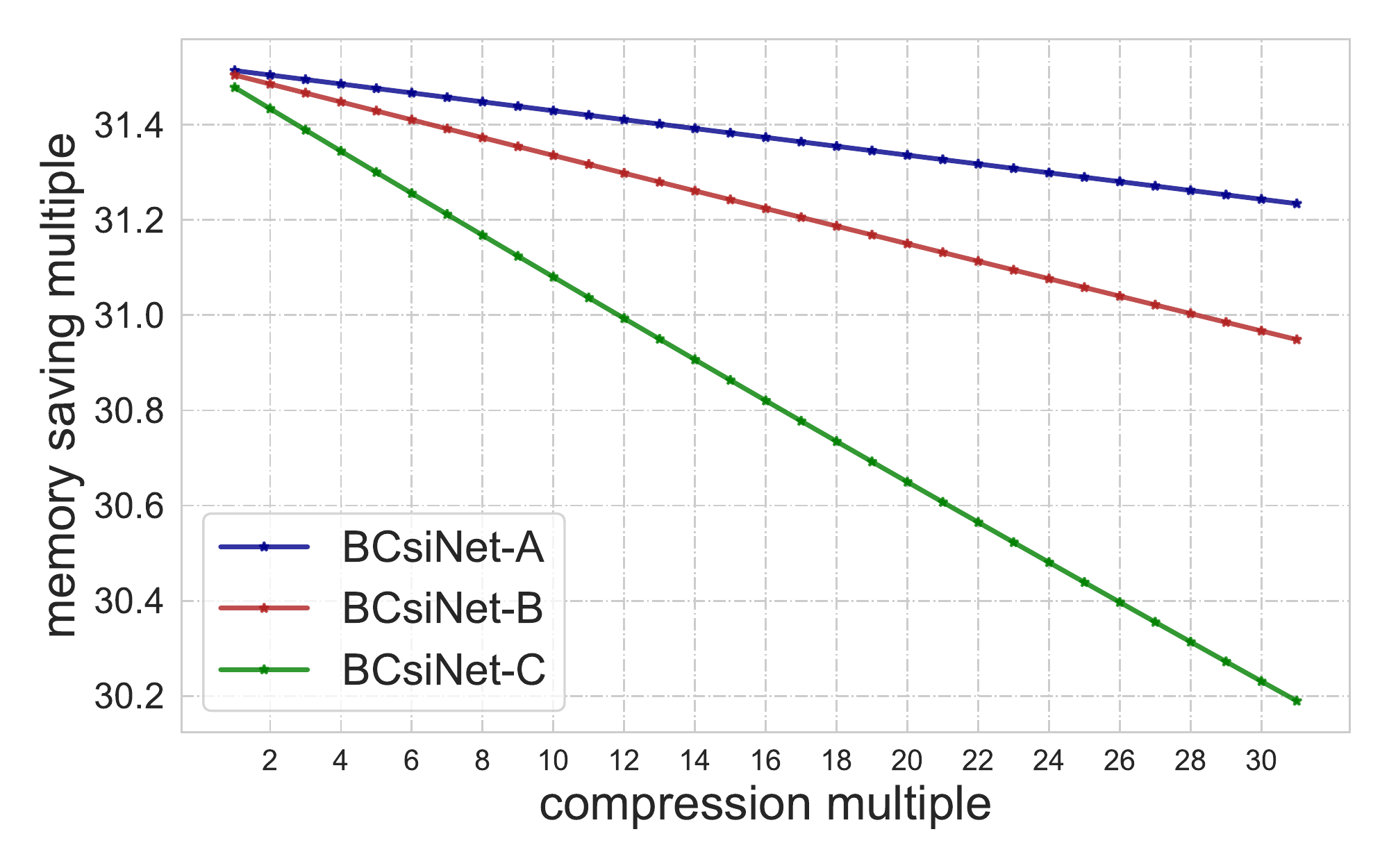}
\caption{Memory saving multiple of BCsiNet variants with head A, B and C over original CsiNet with different compression multiple ($1/\eta$).}
\label{Memory}
\end{figure}

The detailed memory saving multiples of all three heads are depicted in Fig. \ref{Memory}.  It is obvious that the memory saving multiple decays with the increase of compression multiple since FC layer becomes less dominant. However, all three BCsiNet variants achieve over $30\times$ memory saving compared with original CsiNet even when $\eta=1/32$. Therefore, replacing the vanilla head A with head B or C boosts the performance with little cost. Besides, we can see from Table \ref{tab3} and Fig. \ref{Memory} that head B is more cost efficient than head C.

\begin{table}[!b]
\caption{NMSE (dB) and Complexity Comparison between CsiNet and BCsiNet}
\begin{center}
\begin{tabular}{c c|c c | c c}
\Xhline{0.8pt}
\multirow{2}{*}{$\mathbf{\eta}$} & \multirow{2}{*}{\textbf{Methods}} & \multicolumn{2}{c|}{\textbf{complexity at UE}} & \multicolumn{2}{c}{\textbf{NMSE}} \\
    & & mul$^{\mathrm{a}}$  & params & indoor & outdoor \\
\Xhline{0.8pt}
\multirow{3}{*}{1/4} & CsiNet & 1085K & 1049K & -17.36 & -8.75 \\
    & BCsiNet-A2 & 37K & \textbf{33K} & -17.25 & -8.35 \\
    & BCsiNet-B3 &  74K & \textbf{33K} & \textbf{-20.31} & \textbf{-9.77} \\
\hline
\multirow{3}{*}{1/8} & CsiNet & 561K & 525K & -12.70 & \textbf{-7.61} \\
    & BCsiNet-A2 & 37K & \textbf{17K} & -12.38 & -6.26 \\
    & BCsiNet-B3 &  74K & \textbf{17K} & \textbf{-12.77} & -6.86 \\
\hline
\multirow{3}{*}{1/16} & CsiNet & 299K & 262K & -8.65 & -4.51 \\
    & BCsiNet-A2 & 37K & \textbf{8K} & -8.99 & -4.17 \\
    & BCsiNet-B3 & 74K & \textbf{8K} & \textbf{-10.71} & \textbf{-4.52} \\
\hline
\multirow{3}{*}{1/32} & CsiNet & 168K & 131K & -6.24 & \textbf{-2.81} \\
    & BCsiNet-A2 & 37K & \textbf{4K} & -6.79 & -2.69 \\
    & BCsiNet-B3 & 74K & \textbf{4K} & \textbf{-7.93} & -2.74 \\
\Xhline{0.8pt}
\multicolumn{5}{l}{$^{\mathrm{a}}$ ``mul'' refers to the total number of multiplication.} \\
\end{tabular}
\label{tab4}
\end{center}
\end{table}

We denote the vanilla BCsiNet with head A and two RefineNets at BS as ``BCsiNet-A2''. Similarly, BCsiNet with head B and three RefineNets at BS is denoted as ``BCsiNet-B3''. Performance comparison among BCsiNet-A2, BCsiNet-B3 and original CsiNet is demonstrated in Table \ref{tab4}.

Experiments show that the encoder of the proposed BCsiNet at UE is over $30 \times$ lighter than the original CsiNet. Additionally, the inference speed at UE is around $2\times$ faster since the number of multiplication is significantly reduced. Note that the addition complexity remains the same after FC binarization, for which it is omitted in Table \ref{tab3}.

Despite the extremely lightweight encoder, BCsiNet achieves comparable performance over original CsiNet. As a matter of fact, the BCsiNet-B3 scheme outperforms CsiNet with all compression ratio under indoor scenario. For outdoor scenario where channel reconstruction is harder, BCsiNet-B3 dominates only with low compression ratio. Notably, the NMSE performance gap between the lightweight BCsiNet-B3 and the original CsiNet is less that 1dB for high compression ratio under outdoor scenario.

When we focus on the performance of BCsiNet-A2 and BCsiNet-B3, it is apparent that the extra complexity in BCsiNet-B3 strengthens the network capacity. Another interesting observation is that parameters size at UE is almost the same for BCsiNet-A2 and BCsiNet-B3 with arbitrary $\eta$. This further proves the dominance of FC layer at the encoder.

\begin{figure}[t]
\centering
\includegraphics[width=0.48\textwidth]{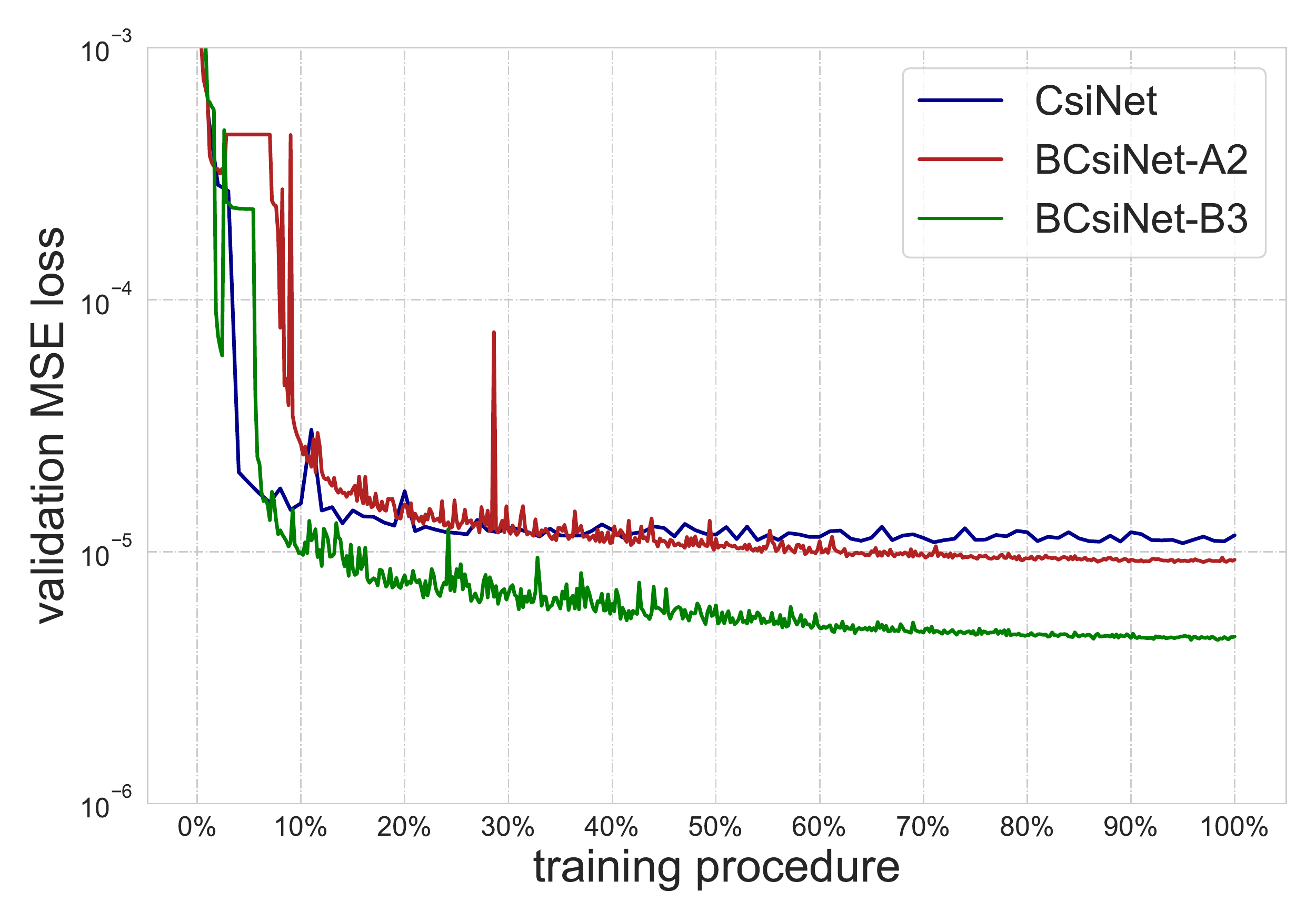}
\caption{Validation loss descending trends of BCsiNet-A2, BCsiNet-B3 and original CsiNet. Experiments are based on in door scenario with compression ratio $\eta=4$.}
\label{Loss}
\end{figure}

Finally, we compare the descending trend of validation loss between BCsiNet and CsiNet in Fig. \ref{Loss}. The proposed BCsiNet converges to a lower MSE loss as expected, and the advantage of BCsiNet-B3 is quite conspicuous. Notably, the training of BCsiNet is less stable at early stage due to the gradient deviation from binarization.

\section{Conclusion} \label{Section-Conclusion}

In this paper, binary neural network technique was applied to CSI feedback task and proved to be effective. After analyzing the complexity bottleneck, a novel feedback network with extremely lightweight encoder named BCsiNet was proposed. Additionally, detailed algorithm of BCsiNet training and inference were designed for better convergency and higher efficiency. Experiments showed that BCsiNet achieved over $30\times$ memory saving and around $2\times$ inference speed up for encoder at UE compared with CsiNet. Moreover, the feedback performance of BCsiNet was comparable with original CsiNet.

\section*{Acknowledgment}
This work was supported in part by the National Key R\&D Program of China under Grant 2017YFE0112300 and Beijing National Research Center for Information Science and Technology under Grant BNR2019RC01014 and BNR2019TD01001.

\bibliographystyle{IEEEtran}
\bibliography{bcsinet.bib}

\end{document}